\documentclass[10pt,conference,letterpaper]{IEEEtran}

\usepackage{amsmath,amsfonts,amsthm,bm}
\usepackage{graphics}
\usepackage{graphicx}
\usepackage[singlelinecheck=false]{caption}
\usepackage{enumitem}
\usepackage[numbers,sort&compress]{natbib}
\usepackage{url}
\usepackage[table,xcdraw]{xcolor}
\usepackage{multirow}
\usepackage{subcaption}
\usepackage{float}
\usepackage{mathtools}
\DeclarePairedDelimiter\ceil{\lceil}{\rceil}
\DeclarePairedDelimiter\floor{\lfloor}{\rfloor}

\begin{document}

\title{SPRITE: A Scalable Privacy-Preserving and Verifiable Collaborative Learning for Industrial IoT}

\author{\IEEEauthorblockN{Jayasree Sengupta}
\IEEEauthorblockA{Indian Institute of Engineering\\ Science and Technology, Shibpur, India\\
Email: jayasree202@gmail.com}
\and
\IEEEauthorblockN{Sushmita Ruj}
\IEEEauthorblockA{University of New South\\Wales, Sydney, Australia\\
Email: sushmita.ruj@unsw.edu.au}
\and
\IEEEauthorblockN{Sipra Das Bit}
\IEEEauthorblockA{Indian Institute of Engineering\\ Science and Technology, Shibpur, India \\ 
Email : sdasbit@yahoo.co.in}}

\maketitle

\begin{abstract}
Recently collaborative learning is widely applied to model sensitive data generated in Industrial IoT (IIoT). It enables a large number of devices to collectively train a global model by collaborating with a server while keeping the datasets on their respective premises. However, existing approaches are limited by high overheads and may also suffer from falsified aggregated results returned by a malicious server. Hence, we propose a Scalable, Privacy-preserving and veRIfiable collaboraTive lEarning (SPRITE) algorithm to train linear and logistic regression models for IIoT. We aim to reduce burden from resource-constrained IIoT devices and trust dependence on cloud by introducing fog as a middleware. SPRITE employs threshold secret sharing to guarantee privacy-preservation and robustness to IIoT device dropout whereas verifiable additive homomorphic secret sharing to ensure verifiability during model aggregation. We prove the security of SPRITE in an honest-but-curious setting where the cloud is untrustworthy. We validate SPRITE to be scalable and lightweight through theoretical overhead analysis and extensive testbed experimentation on an IIoT use-case with two real-world industrial datasets. For a large-scale industrial setup, SPRITE records 65\% and 55\% improved performance over its competitor for linear and logistic regressions respectively while reducing communication overhead for an IIoT device by 90\%.
\end{abstract}

\section{Introduction}

With the growing popularity of Industrial IoT (IIoT), followed by Industry 4.0 which specifically emphasizes on the manufacturing industries \cite{AZH_TII'18}, voluminous amounts of sensitive data are being generated by various smart devices within the industrial floor. To effectively utilise such data in order to provide intelligent customer services and improve the overall quality of service (QoS), industries are inclined towards modelling these data. However, to generate a trained model with high accuracy for better predictive analysis and classification, it is extremely important to utilise training data across different industries or enterprises \cite{Mandal_CCSW'19}. Due to the high sensitivity of these data, industries are unwilling to export them to a centralised server for executing traditional learning algorithms. Thus, collaborative learning \cite{ESORICS'20} has been proposed to enable multiple data owners to construct a global model by leveraging their private data without explicitly sharing them.

Collaborative learning is a promising new paradigm where each device contributes to the global model update by sharing its local model with the cloud/aggregation server without transmitting the private data through the network \cite{arxiv_Choi'20}. In this context, traditional learning algorithms like machine learning (ML) or deep learning can be applied to develop the distributed learning framework. However, deep learning requires an enormous amount of data to begin with in order to search for relevant features and is also extremely intensive in nature because of its layered neural network backbone which are its potential drawbacks. On the contrary, traditional ML algorithms learn from the available features in the dataset to make informed decisions based on the learning which is extremely relevant for structured numerical data generated by applications like healthcare, industries, finance etc. Thus, ML algorithms like regression models have attracted considerable attention over deep learning in such application domain \cite{Wang_IS'21}.

Despite the advantages of collaborative learning, there are two major concerns input data privacy and vulnerability of locally trained models to information leakage. For example, model-inversion attacks \cite{FJR_CCS'15,S&P'19} are able to restore the original training dataset, hence it is essential to keep the local models private. Apart from the privacy issues, the cloud/aggregation server may also behave maliciously by returning forged aggregated results to the devices for making a wrong impact on model update. Under worse circumstances, a server may also return carefully crafted results to the devices for analyzing statistical characteristics of the shared updates and thereby provoke the devices to unintentionally expose sensitive information \cite{Wang_InfoCom'19}. Lastly, since the IIoT devices are typically low-powered, they may become inoperable (i.e. dead) due to energy drainage resulting in device dropout at anytime in the network.

Inspired by the aforementioned challenges, this work focuses on developing a lightweight framework for collaborative learning specific to IIoT in order to ensure data/model privacy and verifiability of the aggregated results returned by the cloud. Our work considers two fundamental regression models: linear and logistic regression and adopts a two-tier fog-based clustered architecture for the collaborative learning setup. The fog-based architecture notably reduces the overheads on the constrained IIoT devices along with minimising the dependence on cloud. Since, in such a setup, the time consuming and computationally intensive local training is typically performed by the low-end IIoT devices, therefore, our proposed training algorithm also handles IIoT device dropout.

\subsection{Related Work}

A brief overview on various privacy-preserving, verifiable distributed machine learning based schemes are discussed.

\noindent \textbf{Privacy-Preservation:} In literature, privacy-preserving aggregation of local models has been achieved either by using differential privacy mechanisms or cryptographic mechanisms. The works \cite{ESORICS'19,KDD'19,TIFS'20,Yunlong_TII'20} concentrating on the former technique usually employ adding noise to the training data or gradients which results in an accuracy drop of the final aggregated model. To handle this issue, certain works \cite{CCS'17,arxiv_Choi'20} have employed a double masking technique along with secret sharing to achieve greater accuracy of the trained model. However, these algorithms are limited by high communication and computation costs. To address the concern of overheads, a group of works have focused on utilizing cryptographic mechanisms like secure multiparty computation (MPC) \cite{Truex'19,CCGRID'20,IoT_Li'21} and homomorphic encryption (HE) \cite{arxiv_Zhu'20,IoT_Zhou'20} which are typically vulnerable to inference threats from third party entities. To address all the aforementioned issues, the latest works \cite{Mandal_CCSW'19,ESORICS'20} employ lightweight secret sharing techniques coupled with additive HE schemes to guarantee privacy-preservation even when majority of participants are corrupted. However, none of these works take into consideration that cloud/aggregation server might behave maliciously and falsify the aggregated final model.

\noindent \textbf{Privacy-Preservation and Verifiability:} To ensure the correctness of the results returned by the server, recent state-of-the-art works \cite{Xianglong_ICC'20,arxiv_Fu'20,Xu_TIFS'20,Han_AIHC'21} have introduced verifiable computation as part of their scheme. However, these works either implement intensive bilinear signatures as part of their verification or use traditional double masking techniques for privacy preservation which has its own pitfalls. Lastly, all of these works are based on either neural networks or deep learning frameworks which are itself complex and computationally intensive. 

\noindent \textbf{Distributed ML Schemes for IIoT:} Recently authors have also proposed privacy preserving ML algorithms \cite{arxiv_Fu'20,arxiv_Liu'20,Lu_TII'20,Zhang_IoT'21} specific to industrial IoT applications. Apart from the intrinsic drawbacks of differential privacy or verifiable computation already discussed above, certain works \cite{Lu_TII'20,Zhang_IoT'21} have introduced blockchain to provide data integrity which results in extra overhead due to the inherent properties of blockchain. Further, all of these works rely largely on IIoT devices to perform certain computations related to blockchain calls or verifiability which is an additional burden for such low-powered devices.

This motivates us to design a privacy-preserving and verifiable collaborative learning algorithm specific to large-scale industries for two fundamental regression models. Our primary focus is to reduce overheads on constrained IIoT devices without sacrificing on the accuracy of the trained model while keeping in mind that such devices may be prone to dropouts. We also address that cloud servers can be malicious and may return forged aggregated results to impact the final model.

%as well as cloud servers can be untrustworthy. 

\subsection{Major Contributions}

\noindent Our contributions can be summarized as below:

\begin{enumerate}[leftmargin=*]
    \item We propose a \textbf{S}calable \textbf{P}rivacy-preserving and ve\textbf{RI}fiable collabora\textbf{T}ive l\textbf{E}arning \textbf{(SPRITE)} algorithm for two fundamental regression models specific to IIoT.
    \begin{enumerate}
        \item Reduces burden on IIoT devices and minimises trust dependence on cloud by introducing fog as a middleware.
        \item Provides robustness to IIoT device dropout by using additive homomorphic based threshold secret sharing.
        \item Ensures verifiability of the aggregated model returned from a malicious cloud (i.e. prevent forgery) by employing verifiable additive homomorphic secret sharing.
    \end{enumerate}
    \item We establish provable security guarantees for the following:
    \begin{enumerate}[leftmargin=*]
        \item privacy-preservation  under an $\textit{honest-but-curious}$ setting and verifiability when cloud is $\textit{untrustworthy}$.
        \item correctness to ensure that the final model is identical to the one produced by the centralised framework.
    \end{enumerate}
    \item Finally, we validate SPRITE with a theoretical overhead analysis accompanied by extensive testbed experiments on an IIoT use-case with two real-world industrial datasets.
    \begin{enumerate}
        \item Establish SPRITE to be scalable and lightweight while maintaining accuracy, data privacy and verifiability.
        \item Justify superior performance over PrivColl \cite{ESORICS'20} with additional security features (e.g. verifiability) where the verification mechanism is lighter compared to VFL \cite{arxiv_Fu'20}.
    \end{enumerate}
\end{enumerate}

\subsection{Organization}

The remaining paper is structured as follows. Section II briefly describes the basic building blocks. Section III discusses the system model. SPRITE is presented in Section IV. Section V explains the security analysis of the scheme. Section VI discusses the experimental analysis with the testbed results. Section VII finally concludes our work.

\section{Preliminaries}

\subsection{Threshold Secret Sharing Scheme \label{tss}}

A \textit{t-out-of-m} threshold secret sharing scheme \cite{Shamir_79} over $\mathbb{F}_q$ splits a value \textit{srt} into $m$ shares in such a way that any $t$ shares can be used to reconstruct $srt$, however a set of at most $(t-1)$ shares makes it impossible to reconstruct $srt$. The scheme consists of the following two algorithms: 

\begin{itemize}[leftmargin=*]
    \item \textbf{\textit{TSS.Share $\bm{(\bm{srt,t,m}) \rightarrow (s_1,s_2, \ldots , s_m)}$:}} Given a secret value $srt$ to be split into $m$ shares and a threshold $t \leq m$, the algorithm outputs a set of shares $(s_1,s_2, \ldots , s_m)$ each belonging to a different entity.
    \item \textbf{\textit{TSS.Reconstruct $\bm{((s_{i_1},s_{i_2}, \ldots , s_{i_t}),m,t) \rightarrow srt}$:}} The algorithm accepts as input any $t$ shares $(s_{i_1},s_{i_2}, \ldots , s_{i_t})$ and outputs the original $srt$.
\end{itemize}

\noindent An additive homomorphic property of the TSS scheme also allows one to compute linear functions of secrets simply by adding the individual shares. 

\subsection{Homomorphic Hash Function \label{hhf}}

A finite field $\mathbb{F}_N$ and a multiplicative group $\mathbb{G}_q$ of prime order $q$ generated by $g$ produces a collision-resistant homomorphic hash function $H:\mathbb{F}_N \rightarrow \mathbb{G}_q$, where $H$ satisfies the following properties \cite{KFM'04}:

\begin{itemize}[leftmargin=*]
    \item \textbf{\textit{One-way:}} computationally hard to compute $H^{-1}(x)$.
    \item \textbf{\textit{Collision-free:}} computationally hard to find $x,y \in \mathbb{F}^N (x \neq y)$, such that $H(x) = H(y)$.
    \item \textbf{\textit{Homomorphism:}} for all $x,y \in \mathbb{F}^N$, it holds $H(x \circ y) = H(x) \circ H(y)$, where ``$\circ$" is either ``+" or ``.".
\end{itemize}

\subsection{Pseudorandom Functions (PRF) \label{prf}}

A family of functions $f:\{0,1\}^n \times \{0,1\}^s \rightarrow \{0,1\}^m$ is said to be a pseudorandom function $\mathcal{F}=\{f:\{0,1\}^n \rightarrow \{0,1\}^m\}$, only if the following conditions are satisfied \cite{TBM'20}:

\begin{itemize}[leftmargin=*]
    \item Given key $K \in \{0,1\}^s$ and input $X \in \{0,1\}^n$, there exists an efficient algorithm to compute $\mathcal{F}_K(X)$.
    \item For an adversary $\mathcal{A}$, there exists a negligible function $\epsilon$, such that:
    $$|{Pr}_{K \leftarrow \{0,1\}^s}[\mathcal{A}^{f_K}]-{Pr}_{f \in \mathcal{F}}[\mathcal{A}^f]|<\epsilon$$
\end{itemize}

\subsection{Verifiable Additive Homomorphic Secret Sharing (VAHSS) using Homomorphic Hash Functions \label{vahss}}

A verifiable additive homomorphic secret sharing scheme is used to compute a final value $y$ corresponding to secret values $x_1, x_2, \ldots, x_n$ over a finite field $\mathbb{F}_N$ along with a proof $\sigma$ that $y$ has been computed correctly \cite{TBM'20}. A homomorphic hash ($H$) based VAHSS has the following algorithms \cite{TBM'20}:

\begin{itemize}[leftmargin=*]
    \item \textbf{\textit{VAHSS.ShareSecret$\bm{(1^\lambda,i,x_i)\rightarrow}$ ($\bm{x_{i1},x_{i2},\ldots,x_{im},}$ $\bm{\tau_i)}$:}} Computes a public value $\tau_i$ corresponding to $x_i$ and then splits $x_i$ into $m$ shares.
    \item \textbf{\textit{VAHSS.PartialEval$\bm{(j,(x_{1j},x_{2j},\ldots,x_{nj}))\rightarrow (y_j)}$:}} Given the $j^{th}$ shares of the secret inputs, the algorithm computes the sum of all such $x_{ij}$ for the given $j,i \in [n]$.
    \item \textbf{\textit{VAHSS.PartialProof$\bm{(j,(x_{1j},x_{2j},\ldots,x_{nj}))\rightarrow(\sigma_j)}$:}} Given the $j^{th}$ shares of the secret inputs, the algorithm outputs the partial proof $\sigma_j=H(y_j)$ respective to $y_j$.
    \item \textbf{\textit{VAHSS.FinalEval$\bm{(y_1,y_2,\ldots,y_m)\rightarrow(y)}$:}} Outputs the summation of all the partial sums $y_1,y_2,\ldots,y_m$.
    \item \textbf{\textit{VAHSS.FinalProof$\bm{(\sigma_1,\sigma_2,\ldots,\sigma_m)\rightarrow(\sigma)}$:}} Computes final proof $\sigma$ as a product of the partial proofs $\sigma_1,\sigma_2,\ldots,\sigma_m$.
    \item \textbf{\textit{VAHSS.Verify$\bm{(\tau_1,\tau_2,\ldots,\tau_n,\sigma,y)\rightarrow(0,1)}$:}} Outputs 1, if $\sigma=\prod_{i=1}^n\tau_i\wedge\prod_{i=1}^n\tau_i=H(y)$ holds or 0 otherwise. 
\end{itemize}

\subsection{Gradient Descent Optimization \label{gdo}}

Gradient Descent is one of the most widely used optimization techniques in machine learning to reach an optimal value for cost function $\mathbf{J}$ \cite{ESORICS'20}. In order to minimize $\mathbf{J}$, the aim is to find the exact values of the coefficient matrix $\theta$ that reduces $\mathbf{J}$ as far as possible. Given a training dataset $x$ with $m$ training samples and label matrix $y$, the cost function is defined as:

\begin{equation}\label{1}
    J(\theta) := \frac{1}{2m} \sum_{i=1}^m(h_\theta(x^{(i)})-y^{(i)})^2
\end{equation}

\noindent Given learning rate $\alpha$, the coefficient matrix $\theta$ is derived as:

\begin{equation}\label{2}
    \theta_j := \theta_j-\alpha \frac{\partial J(\theta)}{\partial \theta_j}
\end{equation}

\noindent Combining equations (\ref{1}) and (\ref{2}) the updated coefficient matrix using gradient descent optimization is as follows:

\begin{equation}\label{3}
    \theta_j := \theta_j-\frac{\alpha}{m}\sum_{i=1}^m(h_\theta(x^{(i)})-y^{(i)})\cdot x_j^{(i)}
\end{equation}

\section{System Model}

This section describes our system model in detail where we propose a two-tier clustered architecture suitable for the collaborative learning setup. Next, we explain the security guarantees and the threat model.

\subsection{Clustered Architecture}

Our model consists of the following major entities:

\begin{itemize}[leftmargin=*]
    \item \textbf{\textit{Industrial IoT (IIoT) Devices$\bm{(D_{m1},\ldots,D_{mn})}$:}} The devices are any form of industrial equipment such as Automated Guided Vehicles (AGVs), conveyor belts, etc. embedded with one or more different types of sensors (e.g. temperature, pressure) to enable a smart infrastructure
    \cite{AZH_TII'18}. Each of these devices has their own datasets including data sensed from the surroundings and/or collected during the production process. Such data may contain private sensitive information and are used for locally training the learning models. IIoT devices choose their nearest fog node and send the locally trained model updates to it for further processing.
    \item \textbf{\textit{Fog Nodes $\bm{(F_1,\ldots,F_m)}$:}} We have introduced the fog layer in our proposed architecture to act as a potential middleware between resource-constrained IIoT devices and the cloud. These nodes are more robust compared to IIoT devices and trustworthy than cloud \cite{TSC'20} and are therefore used for reducing the trust dependence on the cloud \cite{SRB'21}. Smart devices with higher computational power like smartphones/tablets, laptops act as fog nodes in our system. Each fog node $F$ receives locally trained model updates from a set of IIoT devices ($D_{m1},\ldots,D_{mn}$) which form a cluster where $F$ acts as their cluster-head. The fog nodes perform the necessary operations and then offload the rest of the tasks (aggregation, proof of correctness) to the cloud for further processing. On receiving the final aggregated result along with the proof of verification from the cloud, each $F$ verifies the correctness of the received result.
    \item \textbf{\textit{Cloud (Cl):}} The Cloud is responsible for aggregating the inputs received from the fog nodes $F_1,\ldots,F_m$. It also computes a proof of verification which is sent back to the fog nodes along with the final aggregated result.
\end{itemize}

\begin{figure} [!ht]
\begin{center}  
\includegraphics[scale=0.42]{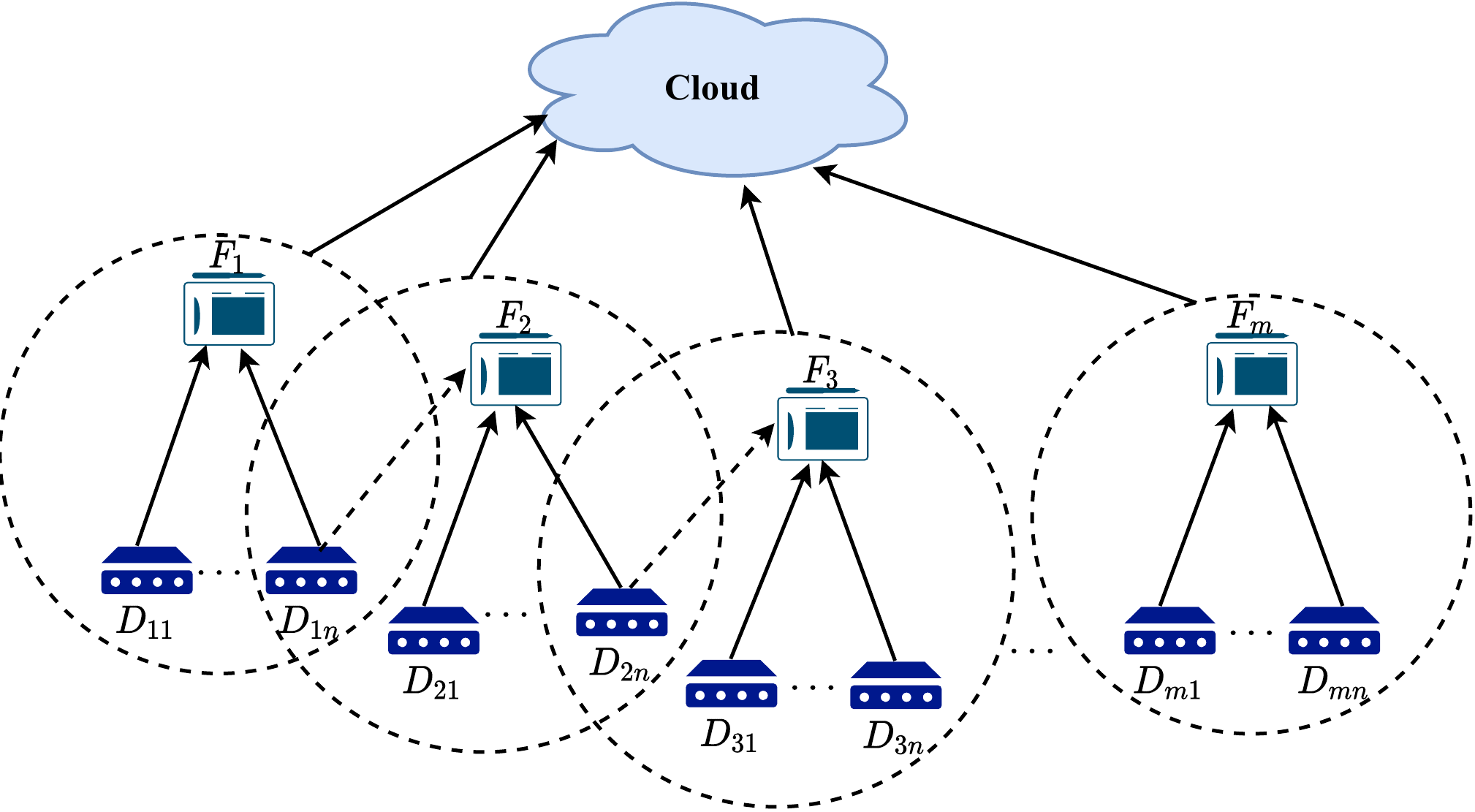}
\setlength{\belowcaptionskip}{-12pt}
\caption{\small \sl Two-tier Clustered Architecture for Collaborative Learning \label{fig:Image1}}
\end{center}  
\end{figure}

\noindent Figure \ref{fig:Image1} shows the two-tier clustered architecture used in our setup. If the nearest fog node to an IIoT device $D_{ij}$ turns offline, $D_{ij}$ can choose any other potential fog node within its communication range.

\subsection{Security Guarantees and Threat Model}

\noindent The design goals, threat models and assumptions are discussed here.

\noindent \textbf{Design Goals:} 

\begin{itemize}[leftmargin=*]
    \item \textbf{\textit{Privacy:}} The privacy of the scheme requires protecting the IIoT device's inputs such as the local datasets and the model parameters. This means that an adversary cannot derive any benign device's local inputs based on its knowledge except with a negligible probability of $\lambda$. No other parties can learn any information about the devices' private inputs.
    \item \textbf{\textit{Correctness:}} For correct inputs from the IIoT devices, the distributed collaborative learning algorithm should deliver the correct model identical to the one produced by the centralised learning algorithm. We claim no correctness of the model if any device uses incorrect inputs.
    \item \textbf{\textit{Verifiability:}} The cloud should only obtain partially evaluated value with its corresponding partial proofs from the fog nodes. The cloud is then supposed to aggregate the values and return a final aggregated result along with a proof of correctness to meet the security requirements of the scheme. The fog nodes should then be able to verify the correctness of the aggregated result received from the cloud.
    \item \textbf{\textit{Efficiency:}} As sensitive data is kept private with the IIoT devices, such devices should perform minimal computation during local training. The rest of the task should be divided between the fog nodes and the cloud in a privacy-preserving and verifiable manner. The computation and communication overheads of each type of component involved in the scheme should be minimal.
\end{itemize}

\noindent\textbf{Threat Model:} In our model, we aim to protect against an \textit{honest-but-curious} adversary $\mathcal{A}$. We consider that a group of IIoT devices and/or Fog nodes compromised by $\mathcal{A}$ follow the protocols correctly but may intend to obtain the intermediate results/information from other nearby devices/nodes. In case of IIoT devices, $\mathcal{A}$ will try to gain information about the local datasets of the other IIoT devices or the local model parameters trained out of them. Similarly, in case of fog nodes, $\mathcal{A}$ will try to gain information about the outputs from other fog nodes. Lastly, we consider the cloud to be \textit{untrustworthy}.

\noindent \textbf{Assumptions:}
The following restrictions have been considered while setting up the assumptions:

\begin{itemize}
    \item We assume that each training iteration is conducted within a fixed time-interval $\mathcal{T}$ in a synchronous way.
    \item It is assumed that all devices use the correct model to compute the local gradient and the model is consistently updated at each round of iteration.
    \item Finally, it is assumed that the IIoT devices/fog nodes never deny to communicate and share part of their secret shares with the corresponding peer nodes.
\end{itemize}

\section{Our Proposed Scheme}

The basic idea of our proposed \textbf{S}calable \textbf{P}rivacy-preserving and ve\textbf{RI}fiable collabora\textbf{T}ive l\textbf{E}arning (\textbf{SPRITE}) scheme is to train a regression model (i.e. linear/logistic regression) in a distributed fashion where the IIoT devices, fog nodes and the cloud collaboratively execute a global gradient computation. 

In SPRITE, Shamir's Secret Sharing \cite{Shamir_79} coupled with additive homomorphism has been used to blind the locally trained model from the fog nodes before sending it out to them. Since \textit{`t'} shares are enough to reconstruct the secret, Shamir's secret sharing ensures robustness to IIoT device dropout in the network up to a certain specified threshold. The novelty of SPRITE lies in delegating the gradient descent optimization (Section \ref{gdo}) algorithm partially to the fog nodes, thereby reducing trust dependence on the cloud. Here, the sensitive partially computed gradients from the fog nodes are split into shares before uploading to the cloud for aggregation thus blinding the original values from it. The cloud in turn provides a proof of correctness along with the final aggregated model update to ensure verifiability of the computed result. Verifiable data aggregation in SPRITE is inspired from VAHSS scheme \cite{TBM'20} as it is lightweight. However, we modify their \textit{n-client, m-server} setup to adopt into our \textit{m-fog, single server} architecture. 

\noindent \textbf{Working Principle:} Each IIoT device holds a local dataset denoted by $\mathcal{D}_{ij}$ and its corresponding model parameter $\theta$ which gets updated in each round of iteration. Each dataset $\mathcal{D}_{ij} \in \mathbb{R}^{m_j \times k}$ is a $m_j \times k$ matrix representing $m_j$ training samples with $k$ features where $m_j$ can vary from device to device. The model parameter $\theta \in \mathbb{R}^{k \times l}$ is a matrix of coefficients, where $l$ is the number of output classes. In our proposed scheme, $M=\sum_{j=1}^N m_j$ (where $N$ is the total number of IIoT devices in the system) and $l$ are publicly known by every participant whereas $k$ is private and is only known to the corresponding data owner $D_{ij}$. Figure \ref{fig:Image2} shows the workflow of SPRITE which consists of the following steps:

\begin{figure} [!ht]
\begin{center}  
\includegraphics[scale=0.47]{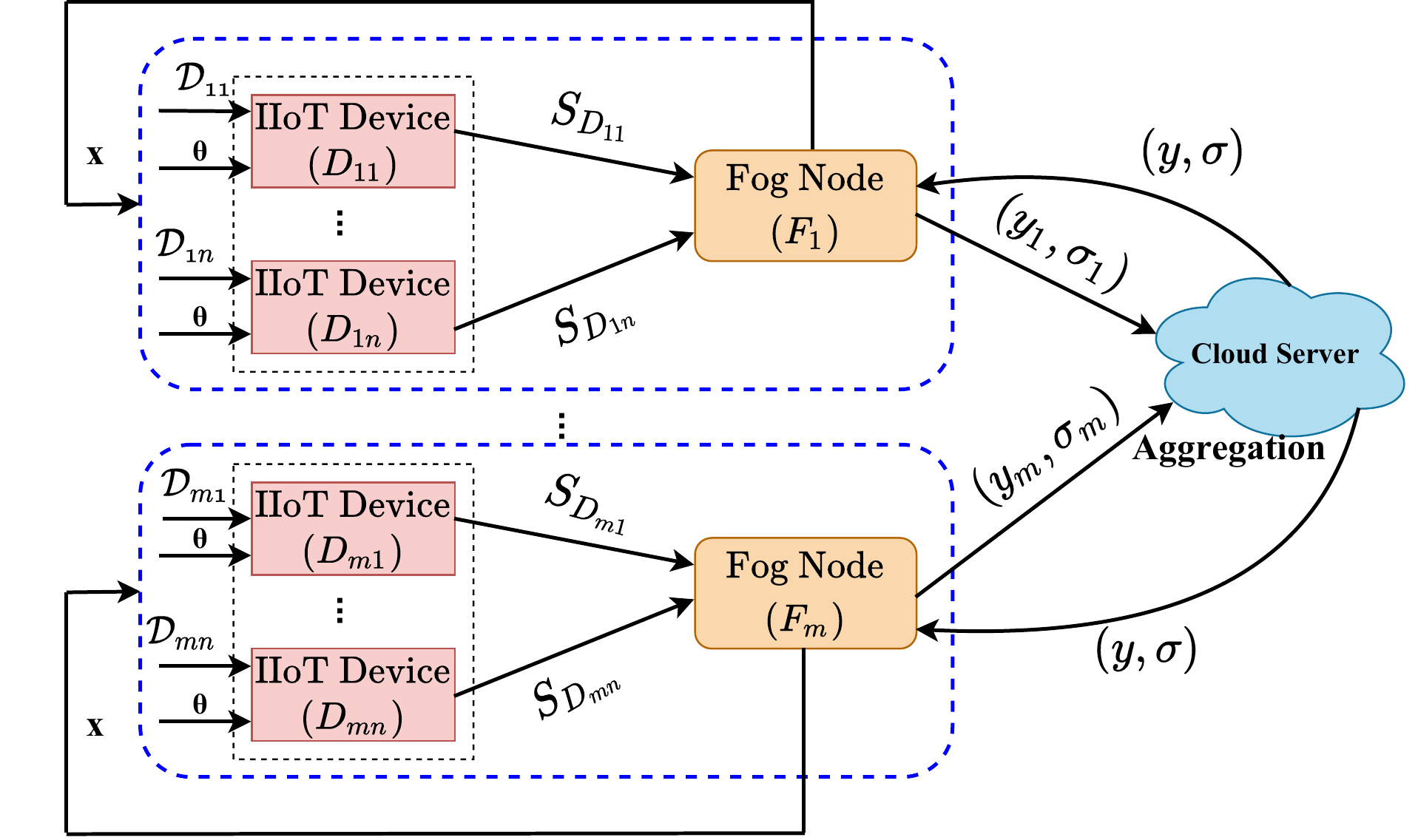}
\setlength{\belowcaptionskip}{-15pt}
\caption{\small \sl Workflow of SPRITE \label{fig:Image2}}
\end{center}  
\end{figure}

\begin{itemize}[leftmargin=*]
    \item \textbf{\textit{System Setup:}} Given the security parameter $\lambda$, a trusted authority (\textit{TA}) generates a secret key $sk_i$ for each fog node $F_i$ and public parameters $pp$. Next, it generates a collision-resistant homomorphic hash function $H:x \rightarrow g^x$ as shown in Section \ref{hhf}. Finally, it publishes a pseudorandom function $\mathcal{F}:\{0,1\}^{l_1} \times \{0,1\}^{l_2} \rightarrow \mathbb{F}$ as per Section \ref{prf}.
    \item \textbf{\textit{Local Training:}} Each IIoT device $D_{ij}$ holds its own training dataset $\mathcal{D}_{ij}$ and the model parameter coefficient matrix $\theta$. In each round of iteration, each $D_{ij}$ multiplies $\mathcal{D}_{ij}$ with $\theta$ to generate $h_{\theta}(\mathcal{D}_{ij}) \in \mathbb{R}^{l \times m_j}$ which is a $l \times m_j$ vector. Next it computes local gradient $g_i$ using $h_{\theta}(\mathcal{D}_{ij})$ over its dataset. Each $D_{ij}$ then splits $g_i$ into $n$ shares using $TSS.Share(g_i,t,n)$ [where $t$ is the threshold number of IIoT devices that need to come together in order to reconstruct the secret value] as explained in Section \ref{tss} and distributes each share to the other corresponding $D_{ij}$. Each $D_{ij}$ then locally adds the shares it receives to produce $S_{D_{ij}}$ and sends this value to the nearest online fog node $F_i$.
    \item \textbf{\textit{Gradient Computation:}} Each fog node $F_i$ receives $S_{D_{i1}},S_{D_{i2}},\ldots,S_{D_{in}}$ values from $n$ IIoT devices lying within its communication range. $F_i$ then computes cumulative gradient $c_i$ using  $TSS.Reconstruct((S_{D_{i1}},S_{D_{i2}},\ldots,S_{D_{in}}),n,t)$. Thus, even if some IIoT devices are offline or busy during this stage and doesn't send its share of $S_{D_{ij}}$, the reconstruction is still feasible given at least $t$ IIoT devices participate. This operation basically computes the summation part of the Gradient Descent algorithm [shown in Eq. (\ref{3})], i.e. $c_i=\sum_{r=1}^{m_j}(h_{\theta}(\mathcal{D}_{ij}^{(r)})-y^{(r)})\cdot \mathcal{D}_{ij}^{(r)}$.
    \item \textbf{\textit{Task Delegation:}} In this stage, each $F_i$ generates the output ${PR}_i = \mathcal{F}_{sk_i}(i,{ts}_i)$ of the pseudorandom function $\mathcal{F}$ [where ${sk}_i \in \{0,1\}^{l_1}$ and ${ts}_i$ is the timestamp associated with $F_i$ such that $(i,{ts}_i) \in \{0,1\}^{l_2}$]. $F_i$ then calls function $VAHSS.ShareSecret(1^\lambda,i,c_i)$ (Section \ref{vahss}) to split $c_i$ into $m$ shares and distribute it to the other corresponding fog nodes. Each $F_i$ also calculate $\tau_i=H(c_i+{PR}_i)$ respective to $c_i$ which is publicly known to the entire system.
    
    $F_i$ then calculates $y_i$ using the corresponding shares of $c_{ij}$ it has received by calling function $VAHSS.PartialEval(i,(c_{1i},c_{2i},\ldots,c_{ni}))$ (Section \ref{vahss}). For the purpose of verification, $F_i$ computes a partial proof $\sigma_i=H(y_i)=g^{y_i}$ using $VAHSS.PartialProof(i,(c_{1i},c_{2i},\ldots,c_{ni}))$ (Section \ref{vahss}). Finally, each $F_i$ sends $(y_i,\sigma_i)$ to the cloud. 
    \item \textbf{\textit{Data Aggregation:}} When the Cloud receives $(y_i,\sigma_i,)$ from the fog nodes $F_1, \ldots, F_m$, it first calls $VAHSS.FinalEval(y_1,y_2,\ldots,y_m)$ to generate $y=y_1+y_2+\ldots+y_m$ and then calls $VAHSS.FinalProof(\sigma_1,\sigma_2,\ldots,\sigma_m)$ as per Section \ref{vahss} to generate final proof $\sigma=\prod_{i=1}^m\sigma_i$. Finally, the cloud sends a 2-tuple packet $(y,\sigma)$ back to each $F_i$.
    \item \textbf{\textit{Result Verification:}} Each $F_i$ on receiving the 2-tuple packet from the cloud, checks whether equation $\sigma=\prod_{i=1}^m\tau_i\wedge\prod_{i=1}^m\tau_i=H(y)$ holds as per $VAHSS.Verify(\tau_1,\tau_2,\ldots,\tau_n,\sigma,y)$ (Section \ref{vahss}). If the equation fails, $F_i$ rejects the result, otherwise, accepts it.
    \item \textbf{\textit{Update Model Parameters:}} On acceptance of the result, $F_i$ computes $x=(\alpha/M)\cdot y$, where $\alpha$ is the learning rate and $M= \sum_{j=1}^N m_j$. $F_i$ then sends back $x$ to each device $D_{ij}$ who updates its respective model parameter as $\theta'=\theta-x$. The updated $\theta'$ is used for the next training iteration.
\end{itemize}

\begin{figure}[!ht]
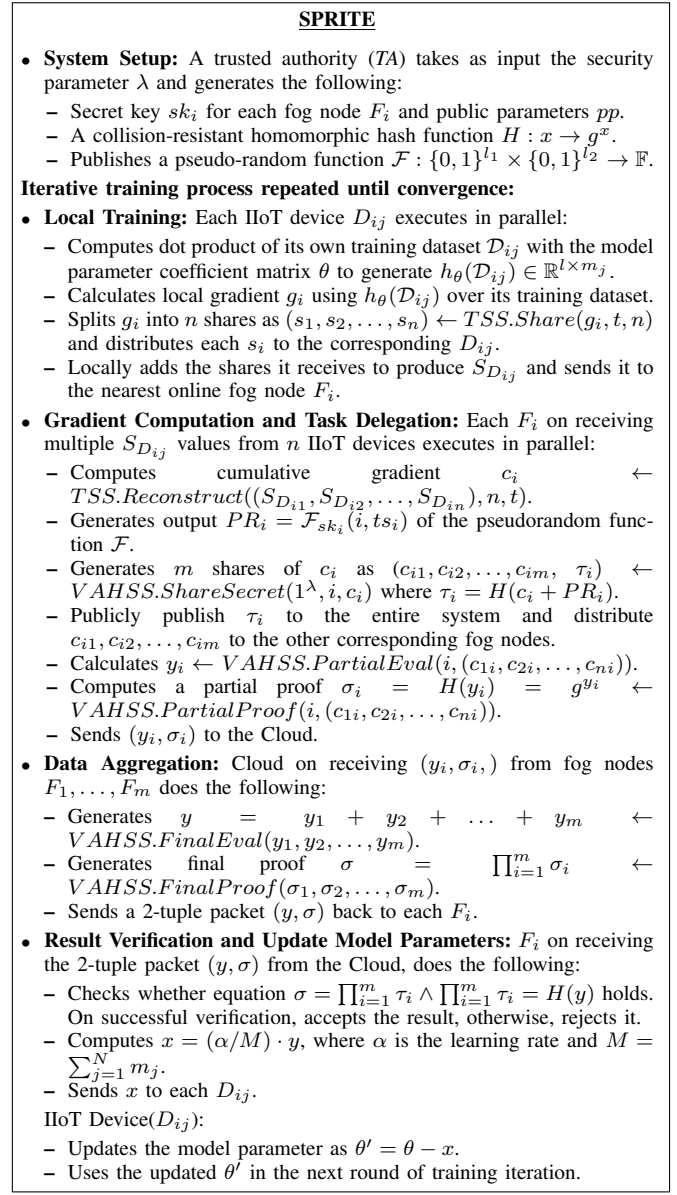

\begin{center}
\fbox{\footnotesize
\begin{minipage}{0.95\columnwidth}
\begin{center}
    \underline{\textbf{SPRITE}}\\
\end{center}
\begin{itemize}[leftmargin=*]
    \item \textbf{System Setup:}
    A trusted authority (\textit{TA}) takes as input the security parameter $\lambda$ and generates the following:
    \begin{itemize}
        \item Secret key $sk_i$ for each fog node $F_i$ and public parameters $pp$.
        \item A collision-resistant homomorphic hash function $H:x \rightarrow g^x$.
        \item Publishes a pseudo-random function $\mathcal{F}:\{0,1\}^{l_1} \times \{0,1\}^{l_2} \rightarrow \mathbb{F}$.
    \end{itemize}
    \end{itemize}
\textbf{Iterative training process repeated until convergence:}
    \begin{itemize}[leftmargin=*]
        \item \textbf{Local Training:} Each IIoT device $D_{ij}$ executes in parallel:
        \begin{itemize}[leftmargin=*]
            \item Computes dot product of its own training dataset $\mathcal{D}_{ij}$ with the model parameter coefficient matrix $\theta$ to generate $h_{\theta}(\mathcal{D}_{ij}) \in \mathbb{R}^{l \times m_j}$.
            \item Calculates local gradient $g_i$ using $h_{\theta}(\mathcal{D}_{ij})$ over its training dataset.
            \item Splits $g_i$ into $n$ shares as $(s_1,s_2, \ldots, s_n) \leftarrow$ $TSS.Share(g_i,t,n)$ and distributes each $s_i$ to the corresponding $D_{ij}$.
            \item Locally adds the shares it receives to produce $S_{D_{ij}}$ and sends it to the nearest online fog node $F_i$.
        \end{itemize}
        \item \textbf{Gradient Computation and Task Delegation:} Each $F_i$ on receiving multiple $S_{D_{ij}}$ values from $n$ IIoT devices executes in parallel:
        \begin{itemize}
            \item Computes cumulative gradient $c_i \leftarrow$  $TSS.Reconstruct((S_{D_{i1}},S_{D_{i2}},\ldots,S_{D_{in}}),n,t)$.
            \item Generates output ${PR}_i = \mathcal{F}_{sk_i}(i,{ts}_i)$ of the pseudorandom function $\mathcal{F}$.
            \item Generates $m$ shares of $c_i$ as ($c_{i1},c_{i2},\ldots,c_{im},$ $\tau_i) \leftarrow$ $VAHSS.ShareSecret(1^\lambda,i,c_i)$ where $\tau_i=H(c_i+{PR}_i)$.
            \item Publicly publish $\tau_i$ to the entire system and distribute $c_{i1},c_{i2},\ldots,c_{im}$ to the other corresponding fog nodes.
            \item Calculates $y_i \leftarrow$ $VAHSS.PartialEval(i,(c_{1i},c_{2i},\ldots,c_{ni}))$.
            \item Computes a partial proof $\sigma_i=H(y_i)=g^{y_i} \leftarrow$ $VAHSS.PartialProof(i,(c_{1i},c_{2i},\ldots,c_{ni}))$.
            \item Sends $(y_i,\sigma_i)$ to the Cloud. 
        \end{itemize}
        \item \textbf{Data Aggregation:} Cloud on receiving $(y_i,\sigma_i,)$ from fog nodes $F_1, \ldots, F_m$ does the following:
        \begin{itemize}[leftmargin=*]
            \item Generates $y=y_1+y_2+\ldots+y_m \leftarrow$ $VAHSS.FinalEval(y_1,y_2,\ldots,y_m)$.
            \item Generates final proof $\sigma=\prod_{i=1}^m\sigma_i \leftarrow$ $VAHSS.FinalProof(\sigma_1,\sigma_2,\ldots,\sigma_m)$.
            \item Sends a 2-tuple packet $(y,\sigma)$ back to each $F_i$.
        \end{itemize}
        \item \textbf{Result Verification and Update Model Parameters:} $F_i$ on receiving the 2-tuple packet $(y,\sigma)$ from the Cloud, does the following:
        \begin{itemize}[leftmargin=*]
            \item Checks whether equation $\sigma=\prod_{i=1}^m\tau_i\wedge\prod_{i=1}^m\tau_i=H(y)$ holds. On successful verification, accepts the result, otherwise, rejects it.
            \item Computes $x=(\alpha/M)\cdot y$, where $\alpha$ is the learning rate and $M=\sum_{j=1}^N m_j$.
            \item Sends $x$ to each $D_{ij}$.
        \end{itemize}
        IIoT Device($D_{ij}$):
        \begin{itemize}[leftmargin=*]
        \item Updates the model parameter as $\theta'=\theta-x$.
            \item Uses the updated $\theta'$ in the next round of training iteration.
        \end{itemize}
    \end{itemize}
\end{minipage}
}
\setlength{\belowcaptionskip}{-10pt}
\caption{\small \sl Detailed description of SPRITE \label{fig:Image3}}
\end{center}  
\end{figure}

The detailed protocol description is illustrated in Fig. \ref{fig:Image3}.

\section{Security Analysis}

This section analyzes the security of SPRITE.

\subsection{Privacy Preservation}

Let us consider the execution of SPRITE where the underlying cryptographic primitives are instantiated with parameter $p$ (corrupted devices) and a set $\mathcal{S}$ of $n$ IIoT devices/Fog nodes (denoted with identities $1, \ldots, n$). At any point during the execution, $\mathcal{S}_i$ is denoted as the subset of devices that have correctly sent their outputs at the $(i-1)^{th}$ training iteration. The input of each device $\mathcal{S}$ is denoted as $x_s$ while the inputs of any subset of devices $\mathcal{S}' \subseteq \mathcal{S}$ is denoted as $x_{\mathcal{S}'}=\{x_s\}_{s \in \mathcal{S}'}$. The following theorem demonstrates that while executing SPRITE, the joint view of any set of less than $t$ (\textit{honest-but-curious}) IIoT devices/Fog nodes, doesn't reveal any information about the other device's inputs.

\noindent \textbf{\textit{Theorem I:}} \textit{There exists a probabilistic polynomial-time (PPT) simulator SIM such that for all $p,t,\mathcal{S}$ with $t \leq |\mathcal{S}|,x_s$, $\mathcal{C}$ where $\mathcal{C} \subseteq \mathcal{S}$, output of SIM is perfectly indistinguishable from output of ${REAL}_\mathcal{C}^{\mathcal{S},t,p}$, given ${REAL}_\mathcal{C}^{\mathcal{S},t,p}(x_\mathcal{S},n)$ is a random variable representing the combined views of parties in $\mathcal{C}$.}

\textit{$${SIM}_\mathcal{C}^{\mathcal{S},t,p}(x_C,n) \equiv {REAL}_\mathcal{C}^{\mathcal{S},t,p}(x_\mathcal{S},n)$$}

\noindent \textit{Proof:} As per \cite{Shamir_79}, we know that threshold $t= \floor*{\frac{n}{2}}+1$ and any coalition of at most $(t-1)$ devices and the server will reveal nothing about the devices' private inputs. The joint view of the parties in $\mathcal{C}$ is independent of the inputs of the parties not in $\mathcal{C}$. Thus, the simulator is able to produce a perfect simulation by running the $\textit{honest-but-curious}$ devices on their true inputs while all other devices on a dummy input by calculating the simulated view of the devices in $\mathcal{C}$. At a granular level, an IIoT device/Fog node splits its gradient $g_i$/$c_i$ into $n/m$ shares and locally adds the shares it receives from its corresponding devices before sending the result $S_{D_{ij}}/y_i$ out onto the next level. Therefore, from the perspective of an adversary $\mathcal{A}$, it is impossible to calculate the gradients $g_i$/$c_i$. Thus, $\mathcal{A}$ will be unable to speculate the original input (local dataset, model parameter, etc.) of the IIoT device/Fog node which is effective in privacy-preservation of the devices' private inputs in a $\textit{honest-but-curious}$ setting.

\subsection{Correctness}

If a centralized gradient descent optimization algorithm $\mathcal{F}_{GD}(\mathcal{D})$ taking dataset $\mathcal{D}$ as input converges to a local/global minima $\eta$, then executing SPRITE with the same hyper parameters (cost function, learning rate etc.) on $
(\mathcal{D}_{11},\ldots,\mathcal{D}_{1n}),\ldots,(\mathcal{D}_{m1},\ldots,\mathcal{D}_{mn})$ [denoted by $\mathcal{F'}_{GD}((D_{ij},\mathcal{D}_{ij}),F_i,Cl)$] also converges to $\eta$.

\noindent \textbf{\textit{Theorem II:}} \textit{SPRITE's gradient descent optimization $\mathcal{F'}_{GD}((D_{ij},\mathcal{D}_{ij}),F_i,Cl)$ algorithm is correct.}

\noindent \textit{Proof:} Let $\theta_j$ denote the model parameters of $\mathcal{F}_{GD}$ at the $i^{th} (i \geq 1) $ training iteration, where $\mathcal{F}_{GD}$ updates $\theta_j$ as follows:
\begin{equation}\label{4}
    \theta_j = \theta_{j-1}-\alpha \frac{\partial J(\theta)}{\partial \theta_j} = \theta_{j-1}-\frac{\alpha}{s}\sum_{i=1}^s(h_\theta(x^{(i)})-y^{(i)})\cdot x_j^{(i)}
\end{equation}

\noindent where $s$ is the number of training samples in dataset $\mathcal{D}$ and $x,y$ are the features and target variable of $\mathcal{D}$ respectively.

\noindent In $\mathcal{F'}_{GD}$, each device $D_{ij}$ computes the following:

$$\Delta \mathcal{D}_{ij}=\sum_{j=1}^{s_i}(h_\theta(x^{(j)})-y^{(j)})$$ where $s_i$ is the number of training sample possessed by $D_{ij}$. After aggregation by fog nodes $F_i$ and cloud $Cl$, $\mathcal{F'}_{GD}$ updates $\theta'_j$ in the $i^{th}$ training iteration as follows: 

\begin{equation}\label{5}
    \theta'_j = \theta'_{j-1}-\frac{\alpha}{M}\sum_{z=1}^M(\Delta \mathcal{D}_{ij})_z
\end{equation}

\noindent where $M=\sum_{i=1}^Ns_i$, given $N$ as the total number of IIoT devices $D_{ij}$ in the setup. Comparing equations (\ref{4}) and (\ref{5}), we find that $\theta_j$ and $\theta'_j$ are updated using the same equations. Therefore, $\mathcal{F'}_{GD}((D_{ij},\mathcal{D}_{ij}),F_i,Cl)$ also converges to $\eta$. Thus,
$\mathcal{F'}_{GD}((D_{ij},\mathcal{D}_{ij}),F_i,Cl)$ $=$ $\mathcal{F}_{GD}(\mathcal{D})$.

\subsection{Verifiability}

\noindent \textbf{\textit{Theorem III:}} \textit{Each Fog node $F_i$ can independently verify the correctness of the aggregated result returned by cloud where the verification scheme can also detect forged results with a very high probability.}

\noindent \textit{Proof:} In order to proof the verifiability of the scheme, we first need to show that $Pr[\textbf{\textit{VAHSS.Verify}}(\tau_1,\tau_2,\ldots,\tau_n,\sigma,y)=1]=1$. By construction, the following holds:

\begin{multline}\label{6}
    y =  \sum_{i=1}^m y_i= \sum_{j=1}^m\sum_{i=1}^m \lambda_{ij} \cdot p_i(e_{ij})= \sum_{i=1}^m\sum_{j=1}^m \lambda_{ij} \cdot p_i(e_{ij})\\=\sum_{i=1}^mp_i(0)=\sum_{i=1}^mc_i
\end{multline}

\noindent where for any $\{i\}_{i=1,\ldots,m}e_{i1},\ldots,e_{im}$ are distinct non-zero field elements and $\lambda_{i1},\ldots,\lambda_{im}$ are corresponding Lagrange's coefficient for any univariate polynomial of degree $t$ over finite field $\mathbb{F}_N$.Additionally, by construction we have the following:

\begin{equation}\label{7}
    \sigma = \prod_{j=1}^m \sigma_j = \prod_{j=1}^m H(y_j)= \prod_{j=1}^m g^{y_j}= g^{\sum_{j=1}^my_j}=g^y=H(y)
\end{equation}

\begin{multline}\label{8}
    \prod_{j=1}^m \tau_j=\prod_{j=1}^m g^{c_j+{PR}_j}=g^{\sum_{j=1}^mc_j}g^{\sum_{j=1}^m{PR}_j}\\=g^{\sum_{j=1}^mc_j}g^{\sum_{j=1}^{m-1}{PR}_j+{PR}_m}=g^{\sum_{j=1}^mc_j}g^{\phi(N)\ceil*{\frac{\sum_{j=1}^m{PR}_j}{\phi(N)}}}\\=g^{\sum_{j=1}^mc_j}=g^{\sum_{j=1}^mc_1+\ldots+c_m}\stackrel{Refer\ Eq.\ (\ref{6})}{=}g^y=H(y)
\end{multline}

\noindent Combining equations (\ref{7}) and (\ref{8}), we get $\sigma=\prod_{i=1}^m\tau_i\wedge\prod_{i=1}^m\tau_i=H(y)$ holds. Thus, \textbf{\textit{VAHSS.Verify}} outputs 1 with probability 1. Now, for an incorrect $y'$, the following happens:

\begin{multline*}
    \textbf{\textit{VAHSS.Verify}}(\tau_1,\tau_2,\ldots,\tau_n,\sigma',y')=1 \Rightarrow \sigma'=\prod_{i=1}^m\tau_i\wedge\prod_{i=1}^m\tau_i\\=H(y') \Rightarrow \prod_{i=1}^m \tau_i = H(y') \stackrel{Refer\ Eq.\ (\ref{8})}{\Rightarrow} H(y) = H(y')
\end{multline*}

\noindent which is a contradiction as $y \neq y'$ and $H$ is collision-resistant. Thus, the probability of the verification scheme executing successfully is less than desirable. Hence, in our proposed scheme $F_i$ can detect forged results with a high probability.

\section{Experimental Analysis and Results}

This section first describes an industrial IoT use case along with the datasets used for experimentation. Next, we evaluate SPRITE both theoretically and experimentally.

\subsection{Industrial IoT Use Case}

%The traditional centralized training algorithms face a lot of challenges like data ownership issues, data privacy etc. and is therefore unsuitable for industrial applications. 
The integration of collaborative learning in IIoT can bring rapid transformation in industrial applications (e.g. smart manufacturing). With the help of collaborative learning manufacturers and suppliers can predict different metrics by accumulating training data from industries distributed across the globe. For example, a company might have its plants located at different geographical locations which are operating under varied climatic conditions. In such a scenario, these plants can jointly train a robust global model while keeping its training data private to the corresponding IIoT devices. This model can then be used to predict the production quantity or product quality depending upon the needs. Similarly, the product suppliers might also want feedback from its clients (e.g. manufacturing industries) and thereby generate a global predictive maintenance model by aggregating local models trained by its various clients. Such a model can then be used to detect faults or predict the remaining useful life of a particular machinery and warn the clients beforehand to take necessary actions and prevent breakdown of assembly lines. The advantages of such a distributed setup are listed below:

\begin{itemize}[leftmargin=*]
    \item Industries can train their models concurrently while sensitive data remains on-premise.
    \item The final global model is able to predict outcomes under different operational conditions. 
\end{itemize}

\subsection{Datasets \label{datasets}}

We have used two different datasets \cite{CCPP,Telemetry} for the experimental evaluation of SPRITE which are discussed below.

\subsubsection{Combined Cycle Power Plant Dataset \label{ccpp}} A combined cycle power plant (CCPP) is a thermodynamic system consisting of one steam turbine, two gas turbines and two heat recovery systems. Correctly predicting the full load electrical power output of a base load power plant is crucial for improving the efficiency and maximising the profit from the available megawatt hours \cite{CCPP'14}. This dataset \cite{CCPP} consists of four input variables: ambient temperature ($AT$), relative humidity ($RH$), atmospheric pressure ($AP$) and exhaust vaccum (V). Full load electrical power output ($P_E$) is the target variable. The dataset is composed of 9568 data points collected over a six-year period (2006–2011) where the CCPP was set to work with a full load over 674 different days. Since, this dataset deals with predicting a real-value, we implement the linear regression version of SPRITE on this dataset.

\subsubsection{Microsoft Azure Telemetry Dataset \label{telemetry}} This real-time telemetry dataset \cite{Telemetry} was created for Predictive Maintenance in 2015 by Microsoft Azure Intelligence Gallery using data simulation methods. The real-time data collected from hundred machines was averaged over every hour to formulate this dataset which is composed of 290601 data points \cite{Telemetry'20}. It consists of thirty input variables like voltage, pressure, rotation, vibration measurements, machine information (e.g. type, age) and historical maintenance records that include failures, error messages. The goal of our experiment is to estimate the probability that a machine will fail due to a component failure in the next 24 hours. The dataset records four components (\textit{comp1, comp2} etc.) which may fail. Thus, we have a total of five classes including \textit{no failure}. Since, this boils down to a classification problem, we implement the logistic regression version of SPRITE on this dataset. As we know that logistic regression can deal with only binary classification problems, we implement a one-vs-rest logistic regression (OVR) for this dataset where the model gets trained once for each class.

\subsection{Theoretical Overhead Analysis \label{TOA}}

The computation and communication overheads are measured in terms of execution time and number of transmitting bytes respectively for a single device of each type. It is assumed that the size of each element computed by any entity is $1$ byte. Table \ref{tab:Table1} summarizes the notations used. We consider $x=4,\ 30$ for linear and logistic regressions respectively and $c=5$ in expressing the overhead measurements. The overheads are calculated for a single round of iteration of the collaborative learning algorithms. Table \ref{tab:Table2} shows the detailed overhead calculation where the competing schemes have been implemented using the datasets described in Section \ref{datasets}. 

\begin{table}[!ht]
\centering
\caption{Notations used for Theoretical Analysis}
\label{tab:Table1}
\scalebox{0.75}{
\begin{tabular}{c|l}
\hline
\rowcolor[HTML]{C0C0C0} 
Notations & \multicolumn{1}{c}{\cellcolor[HTML]{C0C0C0}Meaning} \\ \hline
$T_{Dot}$ & Time to perform a dot product \\ \hline
$T_A/T_S/T_M/T_D$ & Time to perform one addition/subtraction/multiplication/division \\ \hline
$T_E$ & Time to perform an exponentiation \\ \hline
$T_I$ & Time to perform one inverse operation \\ \hline
$T_{Mod}$ & Time to perform a modulus operation \\ \hline
$s$ & Number of training samples possessed by an IIoT device \\ \hline
$x$ & Number of features/columns in the dataset \\ \hline
$n$ & Number of IIoT devices within a cluster \\ \hline
$N/m$ & Total number of IIoT devices/Fog nodes in the system\\ \hline
$\hat{t}$ & Degree of polynomial used in Secret Sharing \\ \hline
$c$ & Number of classes in Logistic Regression \\ \hline
$|H|$ & Message digest produced by hash function $H$ \\ \hline
\end{tabular}%
}
\end{table}

\begin{table*}[!t]
\centering
\caption{Theoretical Overhead Analysis}
\label{tab:Table2}
\scalebox{0.66}{
\begin{tabular}{|c|c|c|c|c|c|c|c|c|c|}
\hline
\multirow{3}{*}{\textbf{Entities}} & \multirow{3}{*}{\textbf{Tasks}} & \multicolumn{4}{c|}{\textbf{Linear Regression}} & \multicolumn{4}{c|}{\textbf{Logistic Regression}} \\ \cline{3-10} 
 &  & \multicolumn{2}{c|}{\textbf{Computation Overhead}} & \multicolumn{2}{c|}{\textbf{\begin{tabular}[c]{@{}c@{}}Communication\\ Overhead (bytes)\end{tabular}}} & \multicolumn{2}{c|}{\textbf{Computation Overhead}} & \multicolumn{2}{c|}{\textbf{\begin{tabular}[c]{@{}c@{}}Communication\\ Overhead (bytes)\end{tabular}}} \\ \cline{3-10} 
 &  & \textbf{SPRITE} & \textbf{PrivColl \cite{ESORICS'20}} & \textbf{SPRITE} & \textbf{PrivColl \cite{ESORICS'20}} & \textbf{SPRITE} & \textbf{PrivColl \cite{ESORICS'20}} & \textbf{SPRITE} & \textbf{PrivColl \cite{ESORICS'20}} \\ \hline
\multirow{4}{*}{\textbf{IIoT Device}} & \textbf{Local Training} & \begin{tabular}[c]{@{}c@{}}$2T_{Dot}+2sT_S$\\ $+(2s-1)T_A$\end{tabular} & \begin{tabular}[c]{@{}c@{}}$2T_{Dot}+2sT_S$\\ $+(2s-1)T_A$\end{tabular} & \multirow{4}{*}{$n(x+1)$} & \multirow{4}{*}{$N(x+1)$} & \begin{tabular}[c]{@{}c@{}}$c[2T_{Dot}+T_D+$\\ $T_A+(x+1)T_E+sT_S]$\end{tabular} & \begin{tabular}[c]{@{}c@{}}$c[2T_{Dot}+T_D+$\\ $T_A+(x+1)T_E+sT_S]$\end{tabular} & \multirow{4}{*}{$cn(x+1)$} & \multirow{4}{*}{$cN(x+1)$} \\ \cline{2-4} \cline{7-8}
 & \textbf{Generating Shares} & \begin{tabular}[c]{@{}c@{}}$(x+1)[(n\hat{t}+n-1)T_A+(n\hat{t}$\\ $+2n-1)T_M+(n-1)T_I]$\end{tabular} & \begin{tabular}[c]{@{}c@{}}$(x+1)[(N-1)(2T_A$\\ $+T_{Mod}+T_S)]$\end{tabular} &  &  & \begin{tabular}[c]{@{}c@{}}$c(x+1)[(n\hat{t}+n-1)T_A+$\\ $(n\hat{t}+2n-1)T_M+(n-1)T_I]$\end{tabular} & \begin{tabular}[c]{@{}c@{}}$c(x+1)[(N-1)(2T_A$\\ $+T_{Mod}+T_S)]$\end{tabular} &  &  \\ \cline{2-4} \cline{7-8}
 & \textbf{Adding Shares} & \begin{tabular}[c]{@{}c@{}}$(x+1)[(n-1)(T_A$\\ $+T_{Mod})]$\end{tabular} & \begin{tabular}[c]{@{}c@{}}$(x+1)[(N-1)(T_A$\\ $+T_{Mod})]$\end{tabular} &  &  & $c(x+1)[(n-1)(T_A+T_{Mod})]$ & \begin{tabular}[c]{@{}c@{}}$c(x+1)[(N-1)(T_A$\\ $+T_{Mod})]$\end{tabular} &  &  \\ \cline{2-4} \cline{7-8}
 & \textbf{Model Update} & $(x+1)T_S$ & $(x+1)(T_S+T_M)$ &  &  & $c(x+1)T_S$ & $c(x+1)(T_S+T_M)$ &  &  \\ \hline
\multirow{6}{*}{\textbf{Fog Node}} & \textbf{Share Reconstruction} & \begin{tabular}[c]{@{}c@{}}$(x+1)[(n\hat{t}+n-1)(T_A$\\ $+T_M)+(n-1)T_{Mod}]$\end{tabular} & \multirow{6}{*}{-} & \multirow{6}{*}{\begin{tabular}[c]{@{}c@{}}$(x+1)(m$\\ $+|H|+2)$\end{tabular}} & \multirow{6}{*}{-} & \begin{tabular}[c]{@{}c@{}}$c(x+1)[(n\hat{t}+n-1)(T_A$\\ $+T_M)+(n-1)T_{Mod}]$\end{tabular} & \multirow{6}{*}{-} & \multirow{6}{*}{\begin{tabular}[c]{@{}c@{}}$c[(x+1)(m$\\ $+|H|+1)]$\end{tabular}} & \multirow{6}{*}{-} \\ \cline{2-3} \cline{7-7}
 & \textbf{Secret Share} & \begin{tabular}[c]{@{}c@{}}$(x+1)[(m\hat{t}+m)T_A+$\\ $(m\hat{t}+m+1)T_E]$\end{tabular} &  &  &  & \begin{tabular}[c]{@{}c@{}}$c(x+1)[(m\hat{t}+m)T_A+$\\ $(m\hat{t}+m+1)T_E]$\end{tabular} &  &  &  \\ \cline{2-3} \cline{7-7}
 & \textbf{Partial Eval} & $(x+1)[(m-1)T_A]$ &  &  &  & $c(x+1)[(m-1)T_A]$ &  &  &  \\ \cline{2-3} \cline{7-7}
 & \textbf{Partial Proof} & $(x+1)[(m-1)T_A+T_E]$ &  &  &  & $c(x+1)[(m-1)T_A+T_E]$ &  &  &  \\ \cline{2-3} \cline{7-7}
 & \textbf{Verify} & $(x+1)[T_D+T_M+T_S]$ &  &  &  & $c(x+1)[T_D+T_M+T_S]$ &  &  &  \\ \cline{2-3} \cline{7-7}
 & \textbf{Partial Derivative} & $2(x+1)T_M$ &  &  &  & $2c(x+1)T_M$ &  &  &  \\ \hline
\multirow{4}{*}{\textbf{Cloud}} & \textbf{Final Eval} & $(x+1)[(m-1)T_A]$ & - & \multirow{4}{*}{\begin{tabular}[c]{@{}c@{}}$(x+1)$\\ $(|H|+1)$\end{tabular}} & \multirow{4}{*}{$(x+1)$} & $c(x+1)[(m-1)T_A]$ & - & \multirow{4}{*}{\begin{tabular}[c]{@{}c@{}}$c[(x+1)$\\ $(|H|+1)]$\end{tabular}} & \multirow{4}{*}{$c(x+1)$} \\ \cline{2-4} \cline{7-8}
 & \textbf{Final Proof} & $(x+1)[(m-1)T_M]$ & - &  &  & $c(x+1)[(m-1)T_M]$ & - &  &  \\ \cline{2-4} \cline{7-8}
 & \textbf{Share Reconstruction} & - & \begin{tabular}[c]{@{}c@{}}$(x+1)[(N-1)(T_A$\\ $+T_{Mod})]$\end{tabular} &  &  & - & \begin{tabular}[c]{@{}c@{}}$c(x+1)[(N-1)(T_A$\\ $+T_{Mod})]$\end{tabular} &  &  \\ \cline{2-4} \cline{7-8}
 & \textbf{Partial Derivative} & - & $(x+1)T_M$ &  &  & - & $c(x+1)T_M$ &  &  \\ \hline
\end{tabular}%
}
\end{table*}

\noindent \textbf{Computation Overhead:} We observe that apparently the computation overhead of SPRITE for IIoT devices is slightly greater compared to PrivColl \cite{ESORICS'20}. This is because we implement Shamir's Threshold Secret Sharing \cite{Shamir_79} contrary to Additive Secret Sharing \cite{BLW_ESORICS'08} used by PrivColl. As a result, at the cost of a little extra computation, we achieve robustness to IIoT device dropout in the system. However, computation overheard for IIoT devices notably reduces  with increasing number of total devices ($N$) in the network, thus making SPRITE fairly scalable. This happens because SPRITE applies secret sharing to IIoT devices ($n$) belonging to a particular cluster contrary to PrivColl which applies secret sharing on the total number of IIoT devices ($N$) in the system. Similarly, the cumulative computation overhead of fog and cloud for SPRITE is greater than that of cloud for PrivColl. This is primarily because we have added a verifiability scheme to ensure that cloud cannot forge/alter its output and secondarily share reconstruction is slightly costlier for our dropout-resilient scheme. 

\noindent \textbf{Communication Overhead:} The focus is to always reduce burden on the low-powered IIoT devices to save its energy. We observe, that SPRITE has been able to substantially reduce the communication overhead for an IIoT device compared to PrivColl. For example, when $(N, n) = (1000,100)$ and $x=4$, communication overhead for an IIoT device is $500$ bytes and $5000$ bytes for SPRITE and PrivColl respectively in case of linear regression. Thus, SPRITE is $10$ times less intensive compared to PrivColl thereby establishing its improved energy management ability over PrivColl. Lastly, we observe that SPRITE ensures additional security (i.e. verifiability) at the cost of a little extra communication by fog and cloud.

\subsection{Testbed Implementation}

Here, we implement SPRITE and validate its performance through an experimental testbed.

\subsubsection{Setup} 

\noindent Experimentation is conducted through testbed where multiple Raspberry Pi(s) (RPI-3B) equipped with DHT11 temperature-humidity sensors act as the Industrial IoT Device (D). A couple of laptops have been used to act as fog nodes and a desktop acts as the server (cloud) in this setup. The specifications of devices used in the testbed are shown in Table \ref{tab:Table3}.

\subsubsection{Evaluation Metrics} We use four metrics to evaluate our prototype such as execution time, Root Mean Square Error (RMSE), R-squared (R2) score and accuracy. Execution time refers to the time taken by an entity (fog, cloud) to perform an operation (e.g. training). RMSE is a standard way to evaluate the error of a model in predicting quantitative data. It is the measure of how well a regression line fits the data points, i.e. how close the observed data points are to the model’s predicted values. On the other hand, R2 is a statistical measure that represents the goodness of fit of a regression model. The ideal value for R2 is 1. Finally, accuracy is defined as the number of correctly predicted data points out of all the data points and is a common metric used for evaluating classification models.

\begin{table}[!htb]
\centering
\caption{Testbed Setup Specifications}
\label{tab:Table3}
\scalebox{0.80}{
\begin{tabular}{|c|c|c|c|}
\hline
\textbf{Specifications} & \textbf{IIoT Device/s (D)} & \textbf{Fog Node/s (F)} & \textbf{Server} \\ \hline
Memory & 1 GBB & 3.8 GiB & 7.7 GiB \\ \hline
Processor & \begin{tabular}[c]{@{}c@{}}Cortex-A53, armv7l\\ @1200MHz * 4\end{tabular} & \begin{tabular}[c]{@{}c@{}}Intel® Core™ i5-7200U\\ CPU @ 2.50GHz * 4\end{tabular} & \begin{tabular}[c]{@{}c@{}}Intel® Core™ i7-6700\\ CPU @ 3.40GHz * 8\end{tabular} \\ \hline
OS & 32-bit Raspbian & \begin{tabular}[c]{@{}c@{}}64-bit Ubuntu\\ 18.04.3\end{tabular} & \begin{tabular}[c]{@{}c@{}}64-bit Ubuntu\\ 18.04.3\end{tabular} \\ \hline
Disk & 16 GB & 455.1 GB & 983.4 GB \\ \hline
\end{tabular}%
}
\end{table}

\begin{table*}[!t]
\begin{minipage}[b]{0.60\linewidth}
\caption{\small \sl Execution Time of each party with increasing number of IIoT devices in the network}
\label{tab:Table4}
\centering
\scalebox{0.63}{
\begin{tabular}{|c|c|c|c|c|c|c|c|c|c|c|}
\hline
\multirow{3}{*}{\textbf{\begin{tabular}[c]{@{}c@{}}(N, n)\\ Devices\end{tabular}}} & \multicolumn{5}{c|}{\textbf{Linear Regression}} & \multicolumn{5}{c|}{\textbf{Logistic Regression}} \\ \cline{2-11} 
 & \multicolumn{3}{c|}{\textbf{SPRITE}} & \multicolumn{2}{c|}{\textbf{PrivColl \cite{ESORICS'20}}} & \multicolumn{3}{c|}{\textbf{SPRITE}} & \multicolumn{2}{c|}{\textbf{PrivColl \cite{ESORICS'20}}} \\ \cline{2-11} 
 & \textbf{\begin{tabular}[c]{@{}c@{}}IIoT Device\\ (sec)\end{tabular}} & \textbf{\begin{tabular}[c]{@{}c@{}}Fog Node\\ (ms)\end{tabular}} & \textbf{\begin{tabular}[c]{@{}c@{}}Cloud\\ (ms)\end{tabular}} & \textbf{\begin{tabular}[c]{@{}c@{}}IIoT Device\\ (sec)\end{tabular}} & \textbf{\begin{tabular}[c]{@{}c@{}}Cloud\\ (ms)\end{tabular}} & \textbf{\begin{tabular}[c]{@{}c@{}}IIoT Device\\ (min)\end{tabular}} & \textbf{\begin{tabular}[c]{@{}c@{}}Fog Node\\ (sec)\end{tabular}} & \textbf{\begin{tabular}[c]{@{}c@{}}Cloud\\ (sec)\end{tabular}} & \textbf{\begin{tabular}[c]{@{}c@{}}IIoT Device\\ (min)\end{tabular}} & \textbf{\begin{tabular}[c]{@{}c@{}}Cloud\\ (sec)\end{tabular}} \\ \hline
(50, 10) & 11.906 & 3270.869 & 33.584 & 18.675 & 160.239 & 10.002 & 97.933 & 1.008 & 11.359 & 4.614 \\ \hline
(100, 10) & 12.280 & 3502.289 & 33.584 & 25.991 & 311.439 & 10.048 & 104.876 & 1.008 & 12.735 & 9.150 \\ \hline
(200, 20) & 13.521 & 4245.284 & 33.584 & 41.569 & 613.914 & 10.173 & 127.166 & 1.008 & 15.790 & 18.225 \\ \hline
(400, 40) & 17.775 & 6453.734 & 33.584 & 73.010 & 1231.909 & 10.577 & 193.419 & 1.008 & 22.313 & 36.763 \\ \hline
(500, 50) & 21.450 & 8187.034 & 33.584 & 89.098 & 1543.854 & 10.916 & 245.418 & 1.008 & 25.535 & 46.123 \\ \hline
(800, 80) & 32.472 & 12766.334 & 33.584 & 142.046 & 2497.809 & 11.898 & 382.797 & 1.008 & 28.567 & 74.742 \\ \hline
(1000, 100) & 45.070 & 19881.294 & 33.584 & 183.745 & 3143.784 & 12.996 & 596.246 & 1.008 & 49.901 & 94.121 \\ \hline
\end{tabular}%
}
\end{minipage}
\hspace{0.3cm}
\begin{minipage}[b]{0.35\linewidth}
\caption{\small \sl Execution Time of each task involved in Verifiability}
\label{tab:Table5}
\centering
\scalebox{0.59}{
\begin{tabular}{|c|cccc|c|}
\hline
\multicolumn{3}{|c|}{\textbf{SPRITE}} & \multicolumn{3}{c|}{\textbf{VFL \cite{arxiv_Fu'20}}} \\ \hline
{\color[HTML]{000000} \textbf{Tasks}} & \multicolumn{1}{c|}{{\color[HTML]{000000} \textbf{\begin{tabular}[c]{@{}c@{}}Linear \\ Regression (ms)\end{tabular}}}} & \multicolumn{1}{c|}{{\color[HTML]{000000} \textbf{\begin{tabular}[c]{@{}c@{}}Logistic \\ Regression (ms)\end{tabular}}}} & \multicolumn{1}{c|}{\textbf{Tasks}} & \textbf{\begin{tabular}[c]{@{}c@{}}p=4\\ (ms)\end{tabular}} & \textbf{\begin{tabular}[c]{@{}c@{}}p=8\\ (ms)\end{tabular}} \\ \hline
{\color[HTML]{000000} Generate Shares} & \multicolumn{1}{c|}{{\color[HTML]{000000} 1.365}} & \multicolumn{1}{c|}{{\color[HTML]{000000} 6.825}} & \multicolumn{1}{c|}{} &  &  \\ \cline{1-3}
{\color[HTML]{000000} Partial Eval} & \multicolumn{1}{c|}{{\color[HTML]{000000} 0.0148}} & \multicolumn{1}{c|}{{\color[HTML]{000000} 0.0740}} & \multicolumn{1}{c|}{\multirow{-2}{*}{Encryption}} & \multirow{-2}{*}{1383} & \multirow{-2}{*}{4916} \\ \hline
{\color[HTML]{000000} Partial Proof} & \multicolumn{1}{c|}{{\color[HTML]{000000} 0.0148}} & \multicolumn{1}{c|}{{\color[HTML]{000000} 0.0740}} & \multicolumn{1}{c|}{} &  &  \\ \cline{1-3}
{\color[HTML]{000000} Final Eval} & \multicolumn{1}{c|}{{\color[HTML]{000000} 0.0044}} & \multicolumn{1}{c|}{{\color[HTML]{000000} 0.0216}} & \multicolumn{1}{c|}{\multirow{-2}{*}{Decryption}} & \multirow{-2}{*}{961} & \multirow{-2}{*}{4379} \\ \hline
{\color[HTML]{000000} Final Proof} & \multicolumn{1}{c|}{{\color[HTML]{000000} 0.0047}} & \multicolumn{1}{c|}{{\color[HTML]{000000} 0.0226}} & \multicolumn{1}{c|}{} &  &  \\ \cline{1-3}
{\color[HTML]{000000} Verify} & \multicolumn{1}{c|}{{\color[HTML]{000000} 0.0172}} & \multicolumn{1}{c|}{{\color[HTML]{000000} 0.0860}} & \multicolumn{1}{c|}{\multirow{-2}{*}{Verification}} & \multirow{-2}{*}{325} & \multirow{-2}{*}{623} \\ \hline
\multicolumn{1}{l}{} & \multicolumn{1}{l}{} & \multicolumn{1}{l}{} & \multicolumn{1}{l}{} & \multicolumn{1}{l}{} & \multicolumn{1}{l}{} \\ \hline
\textbf{Total Time (ms)} & \multicolumn{1}{c|}{1.421} & \multicolumn{1}{c|}{7.103} & \multicolumn{1}{c|}{\textbf{Total Time (ms)}} & 2669 & 9918 \\ \hline
\end{tabular}%
}
\end{minipage}
\end{table*}

\begin{figure*}[!t]
\begin{minipage}[b]{0.60\linewidth}
\centering
\subfloat[Linear Regression]{%
  \includegraphics[clip, width=0.48\linewidth,height=1.2in]{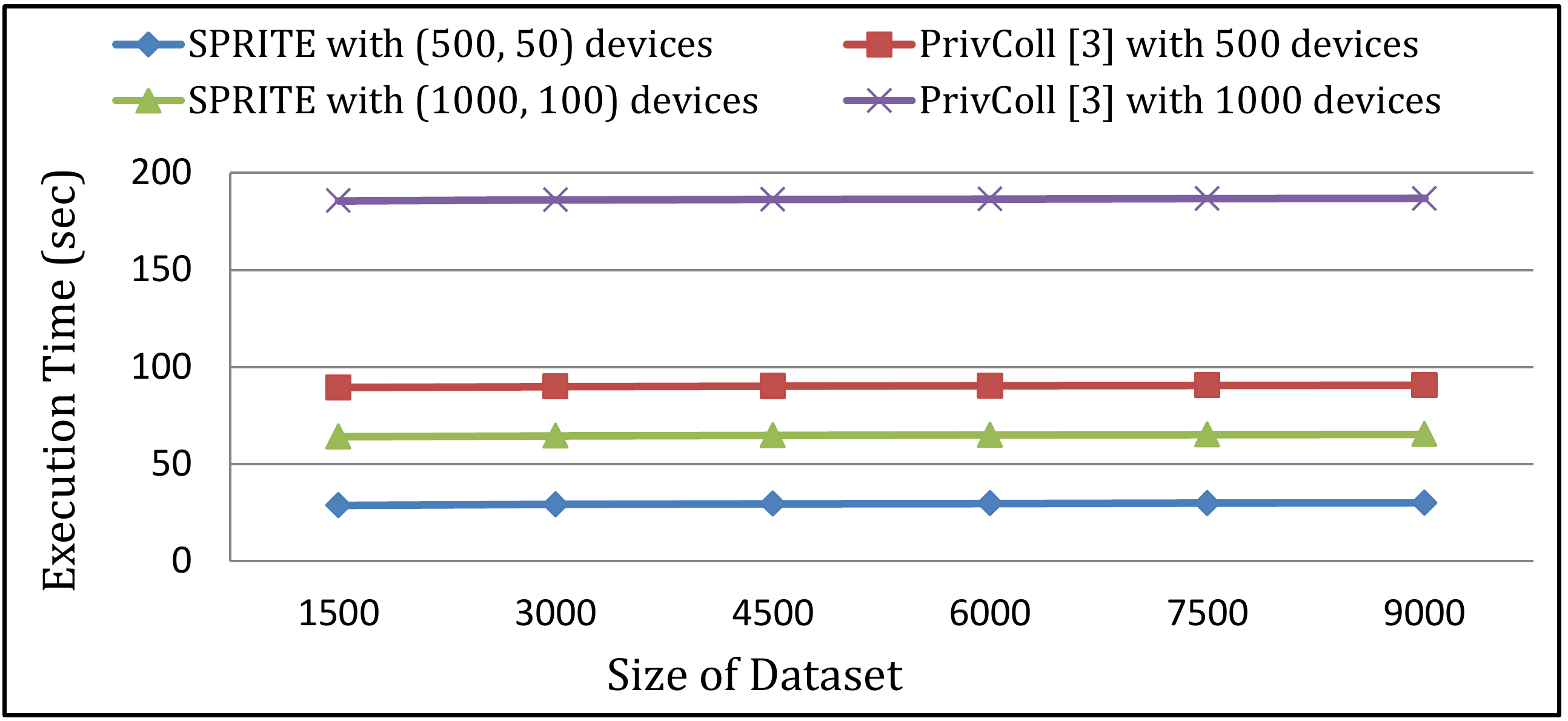}%
  \label{fig:Image4a}
}
\subfloat[Logistic Regression]{%
  \includegraphics[clip, width=0.49\linewidth,height=1.2in]{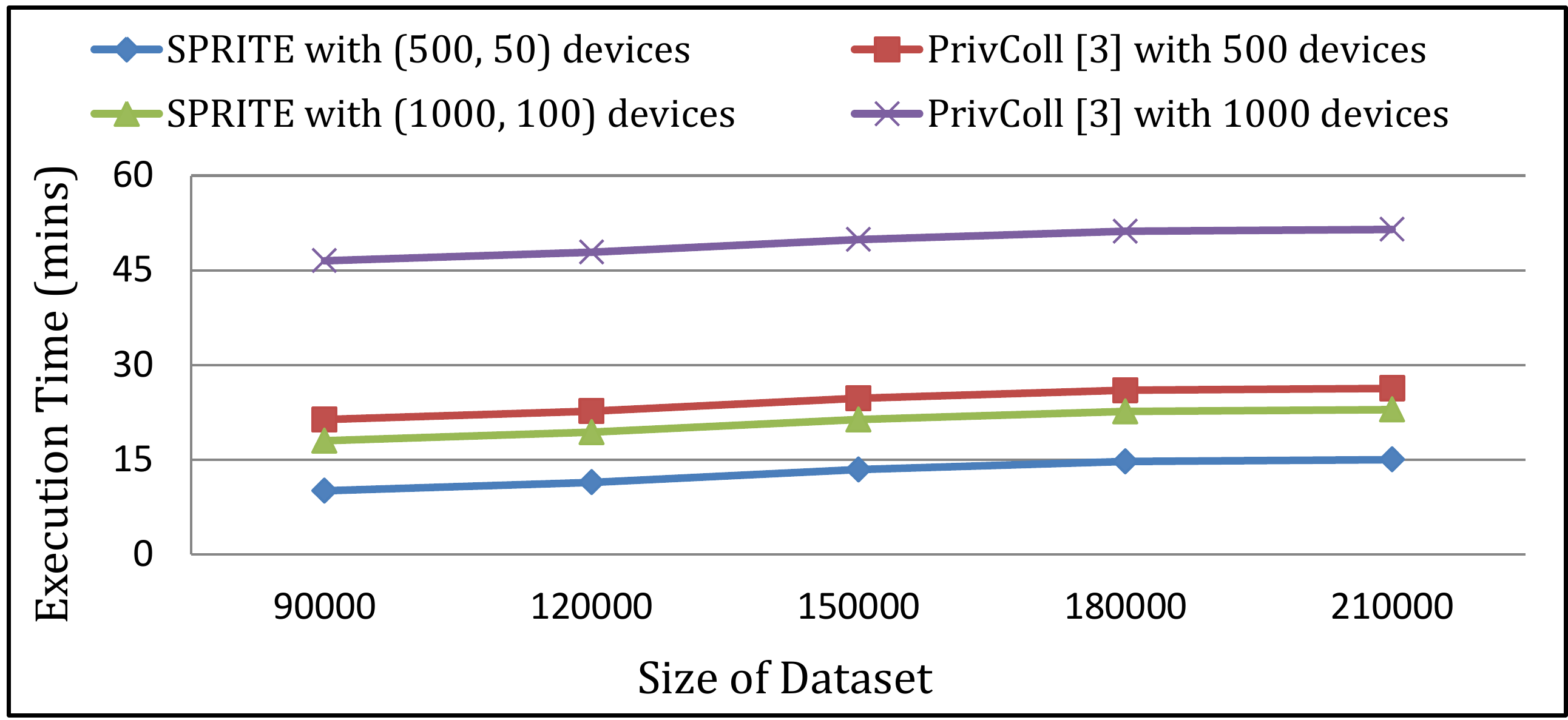}%
  \label{fig:Image4b}
}
\caption{Overall Execution Time with increasing dataset size} \label{fig:Image4}
\end{minipage}
\hspace{0.2cm}
\begin{minipage}[b]{0.36\linewidth}
\centering
\includegraphics[width=\textwidth, height=1.3in]{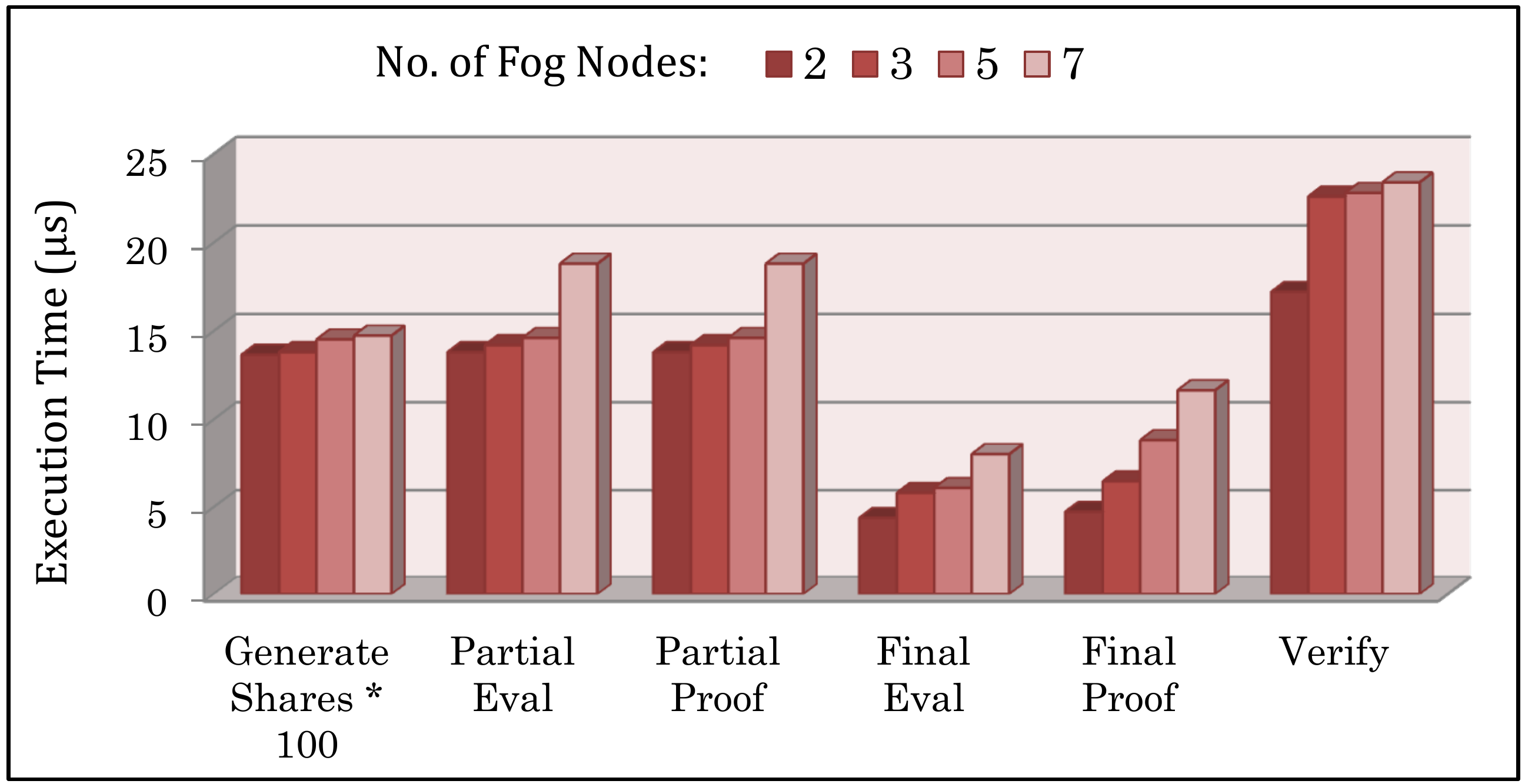}
\caption{\small \sl Execution Time of each sub-task in Verifiability with increasing Fog nodes} 
\label{fig:Image5}
\end{minipage}
\end{figure*}

\subsubsection{Results and Discussion} 

In our experimentation, linear and logistic regression have been applied on the CCPP (Section \ref{ccpp}) and Azure Telemetry (Section \ref{telemetry}) dataset respectively. We conduct five sets of experiments. For each set, the average results of ten independent runs with respect to each task/dataset size has been registered until the training algorithm converged. Among these, in two set of experiments we compare the performance of our proposed scheme with a state-of-the art competitor scheme, PrivColl \cite{ESORICS'20} while in another set of experiment, we compare it with a state-of-the-art verifiability scheme, VFL \cite{arxiv_Fu'20}. The same environment has been used during evaluation of all the competing schemes. We chose the work \cite{ESORICS'20} for comparison with SPRITE because it is the closest competitor to our scheme whereas we chose the work \cite{arxiv_Fu'20} for verifiability comparison because it the most relevant scheme in this context. The rest of the experiments deal only with the performance evaluation of various features specific to SPRITE, hence comparing it with either PrivColl or VFL is irrelevant.

\noindent In the first set of experiment, Table \ref{tab:Table4} shows the execution time of each party while training a global model with the maximum number of available training samples for both the competing schemes. The Table records such results for different combinations of devices ($N, n$) in the network. It is observed that the execution time of IIoT devices is less for SPRITE compared to PrivColl. Moreover, the reduction factor notably increases with growing number of devices in the network. Though the summation of execution times of fog and cloud is more in SPRITE compared to that of just the cloud in PrivColl, the overall execution time of SPRITE is considerably less than PrivColl for both the regression models.

\noindent In the second set of experiment, Table \ref{tab:Table5} shows the execution time of each of the sub-tasks involved in verifiability for SPRITE as well as VFL \cite{arxiv_Fu'20} (where $\textbf{p}$ is the security parameter used by VFL). For our scheme, it is observed that the time taken by each of the sub-tasks is drastically less than the time taken by the sub-tasks in VFL. Even the intensive verification step is significantly faster for SPRITE compared to VFL which is exorbitantly high. Precisely, the total time taken by SPRITE to implement its entire verification mechanism is of the order of a few milliseconds, whereas VFL almost takes a few seconds to implement its verification scheme . Thus, SPRITE's verification mechanism is fairly lightweight compared to VFL.

\noindent In the third set of experiment, we plot (Fig. \ref{fig:Image4a} and \ref{fig:Image4b}) the overall execution time for training a global model for different combination of devices ($N, n$) in the network. We observe from both the figures, that execution time grows marginally with increasing dataset size. We also observe that the execution time of SPRITE is considerably less than PrivColl. Even when training the model with (1000, 100) devices SPRITE records less execution time compared to training the same model with 500 devices using PrivColl. Lastly, we observe that while training the models with (1000, 100) devices SPRITE records 65\% and 55\% improved performance for linear and logistic regressions respectively compared to PrivColl.

\begin{table}[!ht]
\centering
\caption{Various Performance Metrics for Linear and Logistic Regressions}
\label{tab:Table6}
\scalebox{0.80}{
\begin{tabular}{|c|c|c|c|c|}
\hline
\multicolumn{3}{|c|}{\textbf{Linear Regression}} & \multicolumn{2}{c|}{\textbf{Logistic Regression}} \\ \hline
\textbf{Size of Dataset} & \textbf{RMSE} & \textbf{R2} & \textbf{Size of Dataset} & \textbf{Accuracy (\%)} \\ \hline
{\color[HTML]{000000} 1500} & {\color[HTML]{000000} 5.778} & {\color[HTML]{000000} 0.867} & {\color[HTML]{000000} 90000} & {\color[HTML]{000000} 99.63} \\ \hline
{\color[HTML]{000000} 3000} & {\color[HTML]{000000} 4.877} & {\color[HTML]{000000} 0.915} & {\color[HTML]{000000} 120000} & {\color[HTML]{000000} 99.69} \\ \hline
{\color[HTML]{000000} 4500} & {\color[HTML]{000000} 4.803} & {\color[HTML]{000000} 0.925} & {\color[HTML]{000000} 150000} & {\color[HTML]{000000} 99.70} \\ \hline
{\color[HTML]{000000} 6000} & {\color[HTML]{000000} 4.539} & {\color[HTML]{000000} 0.928} & {\color[HTML]{000000} 180000} & {\color[HTML]{000000} 99.74} \\ \hline
{\color[HTML]{000000} 7500} & {\color[HTML]{000000} 4.51} & {\color[HTML]{000000} 0.932} & {\color[HTML]{000000} 210000} & {\color[HTML]{000000} 99.74} \\ \hline
{\color[HTML]{000000} 9000} & {\color[HTML]{000000} 4.486} & {\color[HTML]{000000} 0.933} & \multicolumn{2}{c|}{{\color[HTML]{000000} -}} \\ \hline
\end{tabular}%
}
\end{table}

\noindent In the fourth set of experiment, we evaluate (Table \ref{tab:Table6}) SPRITE in terms of other relevant parameters with varying dataset size. We observe that as the dataset size increases, RMSE score starts decreasing which indicates that the regression line fits better with increasing training samples. Similarly, as the dataset size increases, R2 score starts increasing and reaches a value close to one which is the ideal scenario. Finally, accuracy (\%) also increases with increasing dataset size and reaches a plateau when the model is trained with larger training samples. Thus, we can claim that SPRITE replicates performance similar to the traditional models trained using centralised machine learning algorithms while still achieving the benefits offered by this distributed setup.

\noindent In the last set of experiment, we plot (Fig. \ref{fig:Image5}) execution time of each sub-task involved in verifiability with increasing number of fog nodes to test the scalability of SPRITE. We observe that, execution time of task Generate Shares increases by $\approx$ $1\%-1.5\%$ with increasing fog nodes. The tasks Partial Eval and Partial Proof takes roughly the same time with increase in number of fog nodes, expect for a slight spike when the number of fog nodes is seven. The execution time for tasks Final Eval and Final Proof increases almost linearly with increasing fog nodes. Lastly, execution time of task Verify increases by $\approx$ $0.8\%-2.6\%$ with increasing fog nodes and almost stabilises for larger number of fog nodes, which is a great achievement as it is the most crucial step for any verification scheme.

We summarily observe from the entire testbed experimentation that SPRITE achieves superior performance compared to PrivColl while guaranteeing additional security features. We also observe, SPRITE's verification mechanism is lightweight compared to a state-of-the-art competitor, VFL. It is also observed that SPRITE scales considerably well for a large industrial setup. Lastly, we observe that SPRITE has been able to drastically reduce overheads on the IIoT devices and trust dependence on the cloud by introducing the concept of fog and verifiability. Hence, we claim that SPRITE fits into the requirement of Industrial IoT applications very well.

\subsection{Discussion}

To the best of our knowledge, such a fog-based network design still remains unexplored in an industrial collaborative learning setup. Further, SPRITE is resilient to device-dropout which was missing in PrivColl. Lastly, SPRITE can efficiently handle the issue of forged aggregated model generated by a malicious server, which was a major bottleneck in PrivColl, thereby making SPRITE ideal for real-life applications.

\section{Conclusion}

We have presented SPRITE, a scalable and lightweight privacy-preserving, verifiable collaborative learning framework for IIoT. It is suitable for training both linear and logistic regression models. SPRITE guarantees data/model privacy, verifiability of aggregated results returned by a malicious server thereby preventing forgery on model update. It also ensures robustness to IIoT device dropout. The security of SPRITE is analyzed under an \textit{honest-but-curious} setting where the cloud is \textit{untrustworthy}. SPRITE outperforms the competitors proving its practicality in large-scale industrial applications. In future, we intend to redesign SPRITE to handle collusion cases and detect malicious party/s in case of any corrupt behaviour while adhering to the general workflow.

\bibliographystyle{unsrt}
{\footnotesize
\bibliography{references}}
\end{document}